\documentclass{svproc}
\usepackage{bm,amsmath,amssymb,physics}
\usepackage[dvipdfmx]{graphicx}
\usepackage[dvipdfmx]{color}
\usepackage{multirow}
\def\zI{\mathrm{i}}
\newcommand{\red}[1]{\textcolor[rgb]{0.75,0,0}{#1}}
\newcommand{\blue}[1]{\textcolor[rgb]{0,0,0.8}{#1}}

\begin{document}
\mainmatter              
\title{Time-Dependent Superfluid Band Theory for the Inner Crust of Neutron Stars:\\
Current Status and Future Challenges}
\titlerunning{Time-Dependent Superfluid Band Theory for the Inner Crust}
%
\author{Kazuyuki Sekizawa\inst{1,2,3} \and Kenta Yoshimura\inst{1}}
\authorrunning{Kazuyuki Sekizawa and Kenta Yoshimura} 
%
%
\institute{
Department of Physics, School of Science, Tokyo Institute of Technology, Tokyo 152-8551, Japan
\and
Division of Nuclear Physics, Center for Computational Sciences, University of Tsukuba, Ibaraki 305-8577, Japan
\and
RIKEN Nishina Center, Saitama 351-0198, Japan\\
\email{sekizawa@phys.titech.ac.jp}, \email{yoshimura.k.ak@m.titech.ac.jp}\\WWW home page:
\texttt{https://nuclphystitech.wordpress.com}
}

\maketitle              

\begin{abstract}
In this contribution, current status and future prospects of our ongoing project
are summarized. In the inner crust of neutron stars, a variety of crystalline structures
may emerge, as a result of competition of Coulomb and nuclear interactions, which are
immersed in a sea of superfluid neutrons. The best quantum mechanical approach to study
properties of dripped neutrons under a periodic potential is the band theory of
solids. Concerning the band structure effects on transport properties of neutrons,
however, situation is complicated and there has not been established a
clear consensus yet. To provide a robust conclusion on the band structure effects,
we have developed a fully-microscopic time-dependent band
theory based on time-dependent density functional theory (TDDFT), taking full account
of Fermionic superfluditiy. We have successfully developed a parallel computational
code and applied it to the slab phase of nuclear matter.
We introduce ongoing works and discuss possible future directions.
\end{abstract}


\section{Introduction}

One of the ultimate goals of condensed-matter physics would be to predict
extremely wide-ranging properties of terrestrial matters based on \textit{
ab initio} calculations. Among \textit{ab initio} methods on the market,
the density functional theory (DFT) \cite{DFT} and its time-dependent extension
(TDDFT) \cite{TDDFT1,TDDFT2} are promising approaches, which have shown great
successes in various fields of science, not only condensed-matter physics,
but also quantum chemistry and biology, nuclear physics (see, \textit{e.g.},
Refs.~\cite{Bender(2003),Nakatsukasa(2016),Sekizawa(2019),Colo(2020)}), and so forth. In our ongoing project \cite{Sekizawa(2022),Yoshimura(2023)},
we are aiming at revealing properties of \textit{the densest and the largest
solids in the universe}---the inner crust of neutron stars---microscopically,
applying the band theory of solids \cite{Ashcroft-Mermin} based on nuclear
(TD)DFT for many-nucleon systems\footnote{We note that the nuclear (TD)DFT
to date is not regarded as an \textit{ab initio} approach.}.

\begin{table*}[t]
  \centering
  \caption{
  A summary of the situation concerning the band structure effects in the inner
  crust of neutron stars. An increase of the neutron effective mass ($m_n^\star/m_n>1$),
  which corresponds to entrainment, is highlighted in red; while the opposite trend
  ($m_n^\star/m_n<1$), which we call anti-entrainment, is highlighted in blue.
  In this project, we aim at providing conclusive values of the neutron effective
  mass throughout the inner crust of neutron stars, based on fully self-consistent
  superfluid band theory calculations.
  }\vspace{-2mm}
  \begin{tabular*}{\textwidth}{@{\extracolsep{\fill}}cccccc}
    \hline\hline
    Dimension & Self-consistency & Superfluidity & $m_n^\star/m_n$ & $n_\text{b}$ (fm$^{-3}$) & Ref.      \\
    \hline
    1D        & --         & --         & \red{1.02--1.03}          & 0.074--0.079      & \multirow{2}{*}{\cite{Carter(2005)}} \\
    2D        & --         & --         & \red{1.11--1.40}          & 0.058--0.072      &                                      \\
    3D        & --         & --         & \red{1.07--\textbf{15.4}} & 0.03--0.086       & \cite{Chamel(2005)}                  \\
    3D        & --         & --         & \red{1.21--\textbf{13.6}} & 0.0003--0.08      & \cite{Chamel(2012)}                  \\
    1D        & \checkmark & --         & \blue{0.65--0.75}         & 0.07--0.08$^\dag$ & \cite{Kashiwaba(2019)}               \\
    1D        & \checkmark & --         & \blue{0.59}               & 0.04$^\ddag$      & \cite{Sekizawa(2022)}                \\
    1D        & \checkmark & \checkmark & \blue{0.58}               & 0.07$^\dag$       & Our work \cite{Yoshimura(2023)}      \\
    2D, 3D    & \checkmark & \checkmark & \textbf{??}               & $\lesssim0.07$    & Future works                         \\
    \hline\hline\\[-3mm]
    \multicolumn{4}{l}{\footnotesize $\dag $\,Where appearance of the slab phase is expected.}\\
    \multicolumn{4}{l}{\footnotesize $\ddag$\,With a fixed proton fraction, $Y_p=0.1$.}\\
  \end{tabular*}\vspace{-7mm}
  \label{Table:situation}
\end{table*}

In the context of neutron star studies, the band theory may not yet be regarded
as a popular or a standard approach. While the structure of nuclear ``pasta''
in the inner crust of neutron stars may be well described within ordinary DFT
calculations (see, \textit{e.g.}, Ref.~\cite{Schuetrumpf(2019)}), transport
properties of dripped neutrons may substantially be affected by band structure.
It is because of the Bragg scatterings of dripped neutrons off the periodic potential,
which are expected to \textit{immobilize} part of dripped neutrons, the so-called \textit{entrainment}
effects. In the pioneering works of Refs.~\cite{Carter(2005),Chamel(2005),Chamel(2012)},
it was shown that the effective mass of dripped neutrons is indeed increased
due to the entrainment effects. A notable result was reported for 3D Coulomb
lattices at certain baryon number densities (0.02\,fm$^{-3}\lesssim$\,$n_\text{b}
$\,$\lesssim$\,0.04\,fm$^{-3}$), where the effective mass is predicted to be
more than 10 times larger than the bare neutron mass \cite{Chamel(2005),Chamel(2012)}.
Since such a huge effective mass may potentially affect various interpretations
of astrophysical phenomena of neutron stars, \textit{e.g.}, pulsar glitches
\cite{Andersson(2012),Chamel(2013)glitch,Haskell(2015),Watanabe(2017)} and thermal as well as
crustal properties \cite{Chamel(2009),Chamel(2013),Kobyakov(2013),Kobyakov(2016),Durel(2018)},
we should pay serious attention to the band structure effects in the inner crust
of neutron stars. (See, Ref.~\cite{Chamel(2017)}, for a review of band calculations
of Chamel \textit{et al.} and discussion on related topics.)

Recently, fully self-consistent static \cite{Kashiwaba(2019)} and dynamic \cite{Sekizawa(2022)}
band theories based on nuclear (TD)DFT have been developed and applied for the slab
phase of nuclear matter. In Refs.~\cite{Kashiwaba(2019),Sekizawa(2022)}, it was
shown that the neutron effective mass is actually \textit{reduced} by about 30--40\%,
contrary to the results reported in Ref.~\cite{Carter(2005)}. It has thus been highly
desired to extend the fully self-consistent band theory for 2D and 3D geometries.
One should note that in the aforementioned works pairing correlations (neutron superfluidity)
were neglected. While one may estimate effects of pairing with the Bardeen-Cooper-Schrieffer
(BCS) approximation \cite{Carter(2005)BCS}, on the other hand, a recent work indicates
that a self-consistent [Hartree-Fock-Bogoliubov (HFB) type] treatment is essential to
correctly quantify the effects of pairing on the entrainment effects \cite{Minami(2022)}.
To resolve the arguable situation (see also Table~\ref{Table:situation}), we have recently
extended the theoretical framework for superfluid systems \cite{Yoshimura(2023)}, based
on superfluid (TD)DFT with a local treatment of pairing, known as (time-dependent)
superfluid local density approximation [(TD)SLDA] \cite{TDSLDA1,TDSLDA2,TDSLDA3}.
In this contribution, we provide an overview of our ongoing attempts and discuss
future prospects.

The article is organized as follows.
In Sec.~\ref{Sec:Status}, we briefly summarize the current status of studies
with the fully self-consistent band theory for the inner crust of neutron stars.
In Sec.~\ref{Sec:Future}, we foresee possible future directions and challenges.
In Sec.~\ref{Sec:Summary}, we summarize this contribution.

\vspace{-1mm}
\section{Current Status}\label{Sec:Status}

\vspace{-1mm}
\subsection{Self-consistent band theory and anti-entrainment effects}

The first fully self-consistent band theory calculations based on nuclear
DFT were achieved by Kashiwaba and Nakatsukasa in 2019 \cite{Kashiwaba(2019)}.
In their work, an EDF of the Barcelona-Catania-Paris-Madrid (BCPM) family
was employed, whose form is simpler than Skyrme-type EDFs, especially the
former does not involve the ``microscopic'' effective mass ($m_n^\oplus/m_n=1$).
In the static band calculations, the ``macroscopic'' effective mass of dripped
neutrons, $m_n^\star$, are defined by\footnote{The ``microscopic'' effective mass
refers to the density-dependent one in the EDF, while the ``macroscopic'' one
refers to that associated with the entrainment effects.}
\begin{equation}
m_n^\star/m_n = n_n^\text{f}/n_n^\text{c},
\end{equation}
where $n_n^\text{f}$ is the free (dripped) neutron
number density to which only orbitals with higher single-particle energies than
a potential well of the slab contribute. $n_n^\text{c}=m_n\mathcal{K}_{zz}$
is the conduction neutron number density, where $\mathcal{K}_{zz}$ is the
mobility coefficient which is generally defined as
\begin{equation}
\mathcal{K}_{ij} = 2\sum_\nu\int\frac{\dd^3k}{(2\pi)^3}\frac{\partial^2\varepsilon_{\nu\bm{k}}^{(n)}}{\partial k_i\partial k_j}\theta(\mu_n-\varepsilon_{\nu\bm{k}}^{(n)}).
\end{equation}
Thus, the contribution from each orbital to the diagonal component such as
$\mathcal{K}_{zz}$ is proportional to the curvature of the band. Based on
the static treatment, it was shown \cite{Kashiwaba(2019)} that the macroscopic
effective mass is \textit{smaller} than the bare neutron mass, \textit{i.e.}
$m_n^\star/m_n=0.65$--$0.75$, contrary to the preceding study \cite{Carter(2005)}\footnote{We
note that the discrepancy is partly due to an ambiguity in the definition of
$n_n^\text{f}$, as discussed in Ref.~\cite{Kashiwaba(2019)}.}.

To further deepen our understanding of the entrainment phenomenon, one of
the authors (K.S.) and his collaborators developed a time-dependent version
of the self-consistent band theory based on TDDFT \cite{Sekizawa(2022)}.
In the latter work, an intuitive, dynamical method to quantify the macroscopic
effective mass was proposed. The idea is quite simple. Suppose that we introduce
a constant external electric field, which couples only with protons localized inside
the slabs, that exerts a constant force, $F_\text{ext}$, on protons. Because of the
external force, protons and bound as well as entrained neutrons are expected to be
dragged. According to the classical equation of motion, the collective masses of
the slab, $M_\text{slab}$, and of protons, $M_p$, (per unit area) can be quantified as
\begin{eqnarray}
M_\text{slab}(t) = \dot{P}(t)/a_p(t),\quad
M_p(t) = \dot{P}_p(t)/a_p(t),
\end{eqnarray}
where $a_p(t)$ is the acceleration of center-of-mass position of protons
and $P(t)$ is the total linear momentum, $P(t)=P_n(t)+P_p(t)$, with $P_q(t)
=\hbar\int j_{z,q}(z,t)\dd z$. Since we know the collective masses of the slab
and of protons, the number density of effectively-bound (bound\,+\,entrained)
neutrons can be deduced:
\begin{equation}
n_n^\text{eff.bound} \equiv (M_\text{slab}-M_p)/(am_{n,\text{bg}}^\oplus),
\end{equation}
where $m_{n,\text{bg}}^\oplus$ is the microscopic effective mass for background
neutron number density. Since the conduction neutrons can freely conduct, that is,
they are the counter part of the effectively-bound neutrons, there follows
\begin{equation}
n_n^\text{c} = \bar{n}_n - n_n^\text{eff.bound},
\end{equation}
where $\bar{n}_n$ is the average neutron number density. Based on the real-time
method, it was shown \cite{Sekizawa(2022)} that the macroscopic effective mass
is reduced by 40\%, consistent with Ref.~\cite{Kashiwaba(2019)}. Thanks to the
dynamic simulation, it was demonstrated that the reduction is caused by emergence
of \textit{counter flow} of dripped neutrons, which flows opposite to the external
force. Based on the latter observation, it has been called the \textit{anti-entrainment}
effects \cite{Sekizawa(2022)}.

\subsection{Superfluid extension of the self-consistent band theory}

In this section, we provide the essence of the formulation of the fully
self-consistent (time-dependent) band theory based on (TD)DFT extended
for superfluid systems. We refer readers to Ref.~\cite{Yoshimura(2023)}
for more detailed descriptions.

As the first step, in Ref.~\cite{Yoshimura(2023)}, we have formulated
the fully self-consistent superfluid band theory for the slab phase of
nuclear matter. The key point in formulating the superfluid band theory
is to impose the Bloch's boundary condition to the quasiparticle wave
functions:
\begin{eqnarray}
u_{\nu\bm{k}}^{(q)}(\bm{r}\sigma) &=& \frac{1}{\sqrt{\mathcal{V}}}\widetilde{u}_{\nu\bm{k}}^{(q)}(z\sigma)e^{\zI\bm{k\cdot r}},\label{Eq:qpwf_u_slab}\\
v_{\nu\bm{k}}^{(q)}(\bm{r}\sigma) &=& \frac{1}{\sqrt{\mathcal{V}}}\widetilde{v}_{\nu\bm{k}}^{(q)}(z\sigma)e^{\zI\bm{k\cdot r}},\label{Eq:qpwf_v_slab}
\end{eqnarray}
where $\mathcal{V}$ denotes the volume of the unit cell, $\nu$ is the band index,
$\bm{k}$ is the Bloch wave vector, $q$ is the isospin ($q=n$ for neutrons, $q=p$
for protons) and $\sigma$ is the spin ($\sigma=\uparrow$ or $\downarrow$) of the
quasiparticle wave functions. Taking the direction normal to the slabs as
$z$-direction, the lattice translation vector reads $\bm{T} = T_x\hat{\bm{e}}_x
+ T_y\hat{\bm{e}}_y + na\hat{\bm{e}}_z$, where $T_x$ and $T_y$ are arbitrary real
numbers, $n$ is an arbitrary integer number, $a$ is the slab period (distance
between neighboring slabs), and $\hat{\bm{e}}_i$ ($i=x,y,z$) is the unit vector
along $i$-direction. The dimensionless functions, $\widetilde{u}_{\nu\bm{k}}^{(q)}(z\sigma)$
and $\widetilde{v}_{\nu\bm{k}}^{(q)}(z\sigma)$, are uniform along $x$ and $y$ directions
parallel to the slabs, and manifest the periodicity of the potential, \textit{i.e.},
$\widetilde{u}_{\nu\bm{k}}^{(q)}(z+na,\sigma)=\widetilde{u}_{\nu\bm{k}}^{(q)}(z\sigma)$ and
$\widetilde{v}_{\nu\bm{k}}^{(q)}(z+na,\sigma)=\widetilde{v}_{\nu\bm{k}}^{(q)}(z\sigma)$.

By substituting the quasiparticle wave functions \eqref{Eq:qpwf_u_slab}
and \eqref{Eq:qpwf_v_slab} into the SLDA equation, which is formally similar to the so-called
Hartree-Fock-Bogoliubov (HFB) or Bogoliubov-de Gennes (BdG) equation, we find \cite{Yoshimura(2023)}:
\begin{eqnarray}
\begin{pmatrix}
\hat{h}^{(q)} + \hat{h}_{\bm{k}}^{(q)} - \lambda_q & \Delta_q(z) \\
\Delta_q^*(z) & -\hat{h}^{(q)*} - \hat{h}_{-\bm{k}}^{(q)*} + \lambda_q
\end{pmatrix}
\begin{pmatrix}
\widetilde{u}_{\nu\bm{k}}^{(q)}(z\!\uparrow)\\
\widetilde{v}_{\nu\bm{k}}^{(q)}(z\!\downarrow)
\end{pmatrix}
= E_{\nu\bm{k}}
\begin{pmatrix}
\widetilde{u}_{\nu\bm{k}}^{(q)}(z\!\uparrow)\\
\widetilde{v}_{\nu\bm{k}}^{(q)}(z\!\downarrow)
\end{pmatrix},
\label{Eq:SLDA}
\end{eqnarray}
where $\lambda_q$ is the chemical potential for neutrons $q=n$ and for protons $q=p$.
$\hat{h}^{(q)}(z)$ and $\Delta_q(z)$ are the single-particle Hamiltonian and the pairing field,
respectively, which can be derived from appropriate functional derivatives of an energy density
functional (EDF). In this work, we have developed the formalism with a widely-used Skyrme-type EDF.
The $\bm{k}$-dependent single-particle Hamiltonian is defined by
\begin{equation}
\hat{h}_{\bm{k}}^{(q)}(z) = \frac{\hbar^2\bm{k}^2}{2m_q^\oplus(z)} + \hbar\bm{k\cdot}\hat{\bm{v}}^{(q)}(z),
\end{equation}
where $\hat{\bm{v}}^{(q)}(z)$ is the so-called velocity operator, $\hat{\bm{v}}^{(q)}(z)
= \bigl[ \bm{r}, \hat{h}^{(q)}(z) \bigr]/(\zI\hbar)$.
The $\bm{k}$-dependent term emerges, because of an operation of spatial derivatives on the
Bloch factor $e^{\zI\bm{k\cdot r}}$ in the quasiparticle wave functions \eqref{Eq:qpwf_u_slab}
and \eqref{Eq:qpwf_v_slab}. For the dynamic calculations, we work with the TDSLDA equation
in the velocity gauge \cite{Sekizawa(2022)}. In Ref.~\cite{Yoshimura(2023)}, we have found
that the neutron superfluidity affects little on the entrainment effects, at least for the
slab phase of nuclear matter.

\vspace{-2mm}
\section{Future Directions and Challenges}\label{Sec:Future}

\vspace{-2mm}
\subsection{Extensions to finite temperatures}
\vspace{-1mm}

The extension to finite temperatures can be achieved in a straightforward way.
In (TD)SLDA, finite-temperature effects can be incorporated by modifying
the formulas of various densities \cite{TDSLDA1}. For the number density, for
instance, it reads:
\begin{eqnarray}
n_q(\bm{r},t) &=& \sum_{E_\mu>0}\sum_\sigma
\Bigl[ f(-E_\mu)\bigl|v_\mu^{(q)}(\bm{r}\sigma,t)\bigr|^2 + f(E_\mu)\bigl|u_\mu^{(q)}(\bm{r}\sigma,t)\bigr|^2 \Bigr],
\end{eqnarray}
and similar modifications are applied for other densities as well. The function
$f(E)$ is defined as $f(E)=[1+\exp(\beta E)]^{-1}$ with the inverse temperature,
$\beta=1/(k_\text{B}T)$, where $k_\text{B}$ is the Boltzmann constant. In the superfluid
band theory calculations, the quasiparticle states in the summation, $\mu$, becomes
a combination of quasiparticle states and Bloch wave vectors, $\mu\rightarrow\{\nu,\bm{k}\}$.

\begin{figure}[t]
\begin{center}
\includegraphics[width=\textwidth]{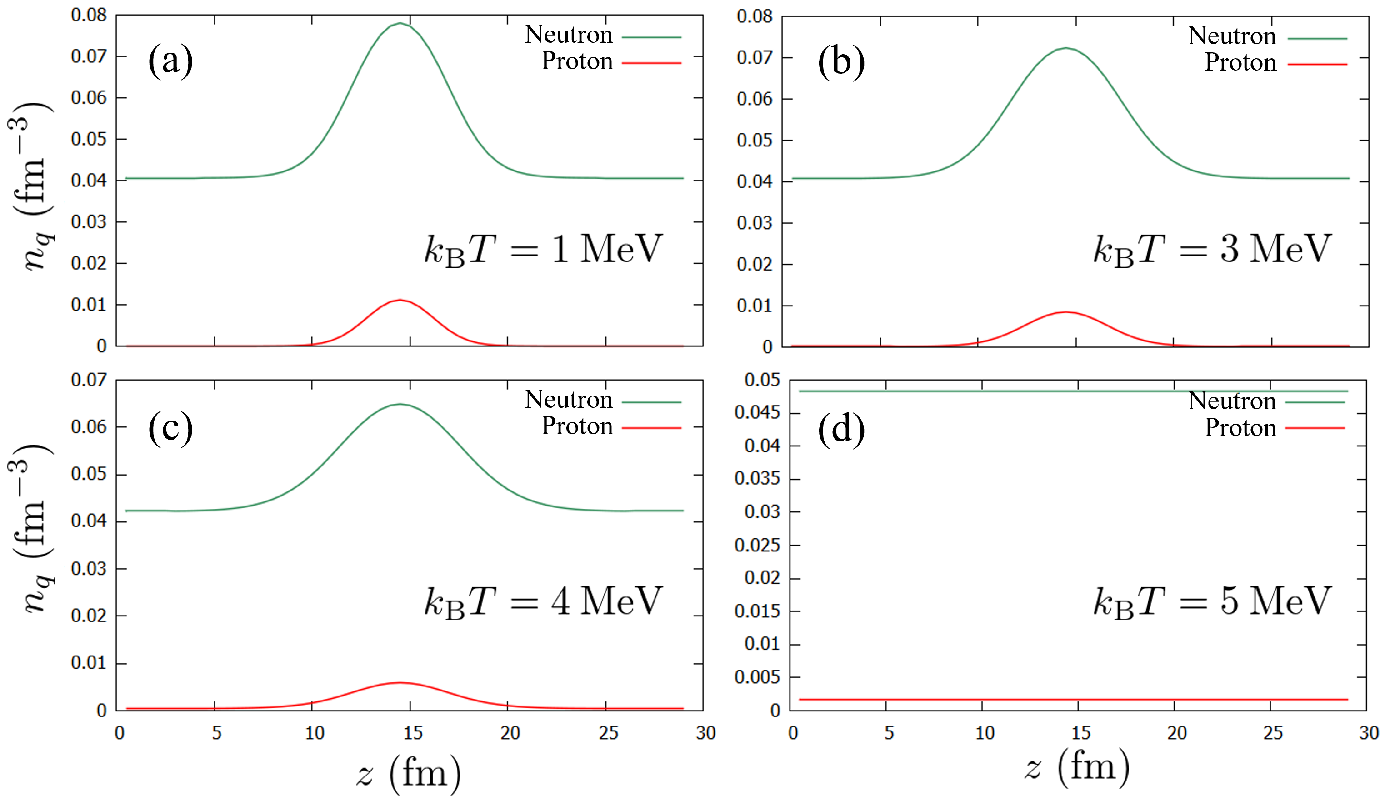}
\end{center}\vspace{-7mm}
\caption{
Neutron (green) and proton (red) density distributions are shown
as a function of $z$ coordinate. In panels (a--d), results of
self-consistent band theory calculations are shown at $k_\text{B}T=1$,
3, 4, and 5\,MeV, respectively.
}\vspace{-5mm}
\label{FIG:FT}
\end{figure}

In Fig.~\ref{FIG:FT}, we show a result of exploratory calculations for $n_\text{b}=0.05$\,fm$^{-3}$
in the $\beta$ equilibrium within the period $a=30$\,fm. In panels (a--d), neutron and
proton density distributions are shown as a function of $z$ coordinate. From panels (a)
to (d), temperature corresponds to $k_\text{B}T=1$, 3, 4, and 5\,MeV, respectively.
From the figure, it is visible that the slab shape becomes more diffusive as temperature
increases. We note that the neutron superfulidity has already gone at $k_\text{B}T\simeq0.8$\,MeV.
At $k_\text{B}T=5$\,MeV, the system eventually becomes uniform, indicating that the nuclear
slab has melt. We aim to predict how nuclear pasta may have been ``cooked'' during/after
a supernova explosion or a neutron-star merger, based on microscopic band theory calculations.
Besides, following the formalism of Ref.~\cite{Schuetrumpf(2020)}, we can now calculate
the elastic and inelastic static structure factors for the slab phase of nuclear matter
within the superfluid band theory at finite temperatures \cite{inprep}. With the latter
quantities we can evaluate neutrino-pasta scattering cross sections and, thus, the
neutrino opacity, which could be an important input for simulations of a supernova
explosion or cooling of a proto-neutron star.

\vspace{-1mm}
\subsection{Extension to finite magnetic fields}
\vspace{-0.5mm}

We can also include finite magnetic-field effects into the description.
The leading effects of a magnetic field comes from the Landau-Rabi quantization
of relativistic electron gas. Because of the formation of Landau levels,
the electron fraction is enhanced, which also enhances the proton fraction
through the charge neutrality condition. In this way, the magnetic field
can directly alter the composition of ($n,p,e$) matter and affects the
structure of the crusts (see, \textit{e.g.}, Ref.~\cite{Sekizawa(2023)},
and references therein).

The magnetic field can also interact with nucleons through the magnetic moments.
On one hand, however, we know that a really strong magnetic field, as large as
$B\gtrsim10^{17}$\,G, is required to affect the nuclear structure (see, \textit{e.g.},
Refs.~\cite{Arteaga(2011),Stein(2016)}). It is simply because of the fact that such
a strong magnetic field is necessary to make the energy shifts caused by a magnetic
field on the order of MeV. Thus, the outer crust of the majority of neutron stars
(with $B\simeq10^{12}\text{--}10^{13}$\,G or, at most, $\approx10^{16}$\,G in
strongly-magnetized magnetars) magnetic field effects on nuclear structure could
safely be neglected. On the other hand, in the inner crust of neutron stars,
dripped neutrons exhibit complex band structure, where band gaps can be much
smaller than 1\,MeV. It may be possible that a magnetic field alters the band
structure even for a smaller magnetic-field strength ($\lesssim10^{14\text{--}15}$\,G).
Thus, we are going to explore possible effects of a magnetic field on crustal
and/or transport properties of dripped neutrons in the inner crust of neutron stars.

The extension to include the finite magnetic-field effects are in progress.
By allowing \textit{spin polarization} of neutrons and protons within the asymmetric
SLDA (ASLDA) framework \cite{TDSLDA1,ASLDA}, we plan to investigate spin polarization
of nuclear matter under a superstrong magnetic field in a fully microscopic framework.
At the same time, it would be interesting to explore possible emergence of the
\textit{spin-triplet} pairing under such an extreme condition (\textit{cf.}
Ref.~\cite{Tajima(2023)}). We consider that it is possible to extend our framework
to describe the spin-triplet pairing by introducing the spin-current pair density,
as recently discussed in Ref.~\cite{Hinohara(2023)}.

\vspace{-1mm}
\subsection{Extension to higher spatial dimensions}

We anticipate that the extension to higher spatial dimensions (\textit{i.e.},
2D and 3D geometries) is not only straightforward, but also many knowledges
and techniques can be brought from the ordinary band theory for terrestrial
solids in condensed-matter physics. We thus consider that the main obstacle
to achieve this extension will be the computational cost. For the slab phase
(a 1D crystal) with a typical numerical setup for actual simulations (with
$\Delta z=0.5$\,fm, $a\simeq30$\,fm), the number of quasiparticle wave
functions is estimated to be 2,880,000 \cite{Yoshimura(2023)}. That is,
we have already dealt with time evolution of \textit{millions} of
quasiparticle wave functions by solving complex, non-linear, coupled
partial differential equations (PDEs). Even though the computational
cost per quasiparticle wave function is rather small for the 1D system,
we have implemented the MPI parallelization for $k_z$ to reduce actual
computation time.

For the rod phase (a 2D crystal), for instance, taking the direction
of rod-shaped nuclear clusters along $z$ direction, the number of
quasiparticle wave functions becomes $N_x\times N_y\times N_{k_x}\times
N_{k_y}\times N_{k_\parallel}\times \text{2 ($n$ and $p$)}\times \text{2 ($u$ and $v$)}$,
where $k_\parallel$ represents the Bloch wave number along the rods.
We may, in this case, increase the lattice spacing to $\Delta z=1.0$
or 1.2\,fm, leading to $N_x\times N_y\simeq 625\text{--}900$, assuming
the area of a unit cell is roughly $30^2$\,fm$^2$. Those lattice spacings
are often used for nuclear (TD)DFT calculations. By roughly estimating
$N_{k_x}\times N_{k_y}\times N_{k_\parallel}\approx80^3$, the total
number of quasiparticle wave functions already exceeds 1 billion!

We note that the estimated numbers of quasiparticle wave functions will be doubled,
if we introduce an interaction that mixes spin states (\textit{e.g.}, the spin-orbit
interaction) or an external field that breaks the time-reversal symmetry (\textit{e.g.},
a magnetic field). Although one could reduce significantly the number of quasiparticle
wave functions by employing an EDF that does not involve the microscopic (density-dependent)
effective mass (\textit{i.e.}, $m_n^\oplus/m_n=1$), as was employed in Ref.~\cite{Kashiwaba(2019)},
we are facing a computational challenge to solve an enormous number of complex, non-linear,
coupled PDEs for time evolution of quasiparticle wave functions to simulate dynamics
of neutron star crustal matter within a fully self-consistent superfluid band theory.
We plan to use the GPGPU parallelization, successfully employed in Refs.~\cite{TDSLDA3,COCG},
to attack the problem.

\vspace{-2.5mm}
\section{Summary}\label{Sec:Summary}
\vspace{-1.5mm}

In this project, we aim at establishing a microscopic, fully self-consistent
time-dependent band theory taking into account various effects such as superfluidity,
finite temperatures and magnetic fields, and so on. We have just put forward the first
step towards this goal in Ref.~\cite{Yoshimura(2023)}. We note that the present formalism
can be used for a unified description of a neutron star, although the band structure
effects are absent in the outer crust and core regions. If the project completes
successfully, it will provide us a fully microscopic equation of states (EoS) with
effects of band structure and neutron superfluidity as well as proton superconductivity
at arbitrary temperatures and magnetic fields, which will be the most reliable class of
EoS on the market that can be used for a variety of applications such as simulations of
supernova explosions or neutron star mergers. The next step, the finite-temperature
extension, will be published elsewhere \cite{inprep}.

\vspace{-1mm}
\paragraph{Acknowledgments.}
Meetings in the A3 Foresight Program supported by JSPS are acknowledged for useful discussions.
One of the authors (K.Y.) would like to acknowledge the support from the Hiki Foundation, Tokyo Institute of Technology.
This work mainly used computational resources of the Yukawa-21 supercomputer at Yukawa Institute
for Theoretical Physics (YITP), Kyoto University. This work also used (in part) computational resources
of the HPCI system (Grand Chariot) provided by Information Initiative Center (IIC), Hokkaido University,
through the HPCI System Project (Project ID: hp230180). This work is supported by JSPS Grant-in-Aid for
Scientific Research, Grant Nos. 23K03410 and 23H01167.

%
%

\end{document}